\documentclass[prd,aps,twocolumn,nofootinbib,superscriptaddress,floatfix,preprintnumbers]{revtex4-1}
\usepackage{amsmath, float, graphicx,amssymb,multirow,pgfplots,pgfplotstable}
\usepackage{tikz}
\usepackage[colorlinks=true,breaklinks=true,citecolor=blue,linkcolor=blue]{hyperref}
\usepackage{ulem}
\usepackage{comment}
\usepackage[caption=false]{subfig}
\usetikzlibrary{shapes.geometric,arrows}
\usepackage{cleveref}
\usepackage{slashed}

\newcommand\barparena[1]{\overset{%
   \scriptscriptstyle(-)}{#1}}

\begin{document}
\preprint{CETUP-2024-005}
\title{New Laboratory Constraints on Neutrinophilic Mediators}

\author{P.~S.~Bhupal Dev}
\email{bdev@wustl.edu}
\affiliation{Department of Physics and McDonnell Center for the Space Sciences, Washington University, St.~Louis, MO 63130, USA}

\author{Doojin Kim}
\email{doojin.kim@usd.edu}
\affiliation{Mitchell Institute for Fundamental Physics and Astronomy, Department of Physics and Astronomy, Texas A\&M University, College Station, TX 77843, USA}
\affiliation{Department of Physics, University of South Dakota, Vermillion, SD 57069, USA}

\author{Deepak Sathyan}
\email{dsathyan@tamu.edu}
\affiliation{Mitchell Institute for Fundamental Physics and Astronomy, Department of Physics and Astronomy, Texas A\&M University, College Station, TX 77843, USA}
\affiliation{Maryland Center for Fundamental Physics, Department of Physics, University of Maryland, College Park, MD 20742, USA}

\author{Kuver Sinha}
\email{kuver.sinha@ou.edu}
\affiliation{Department of Physics and Astronomy, University of Oklahoma, Norman, OK 73019, USA}

\author{Yongchao Zhang}
\email{zhangyongchao@seu.edu.cn}
\affiliation{School of Physics, Southeast University, Nanjing 211189, China}

\begin{abstract}
Neutrinophilic mediators are well-motivated messenger particles that can probe some of the least known sectors of fundamental physics involving nonstandard interactions of neutrinos with themselves and potentially with dark matter. In particular, light mediators coupling to the active neutrinos will induce new decay modes of the Standard Model mesons (e.g., $\pi^\pm, K^\pm \to \ell^\pm + \barparena{\nu} + \phi$), charged leptons (e.g., $\tau^\pm \to \pi^\pm + \barparena{\nu} + \phi$), and gauge bosons (e.g., $Z \to \nu + \bar\nu + \phi$). A common lore is that these decays suffer from infrared divergences in the limit of the vanishing mediator mass, i.e., $m_\phi \to 0$. Here, we show that including the 1-loop contributions of these mediators to the standard 2-body decays (e.g., $\pi^\pm,\,K^\pm \to \ell^\pm + \barparena{\nu}$, etc.), the infrared divergence from the 3-body decay cancels out exactly by virtue of the Kinoshita–Lee–Nauenberg theorem. Including these cancellation effects, we then update the existing laboratory constraints on neutrinophilic scalar mediators, thereby extending the limits far beyond the decaying parent particle mass and excluding a wider range of parameter space.
These new ``physical'' limits derived here have significant implications for the future detection prospects of nonstandard neutrino (self-)interactions.
\end{abstract}

\maketitle


\section{Introduction}

Neutrinos are the least understood out of the Standard Model (SM) particles. In particular, they can have potentially large nonstandard interactions and can serve as a portal to beyond the SM (BSM) physics. While nonstandard neutrino interactions with charged SM fermions are readily probed with neutrino scattering and oscillation experiments~\cite{Farzan:2017xzy, Proceedings:2019qno}, neutrino self-interactions~\cite{Berryman:2022hds} and possible connections to dark matter~\cite{Blennow:2019fhy} can be effectively probed by studying the interactions of neutrinophilic mediators. In fact, it is common to have (light) scalar or vector bosons mediating the self-interactions among active neutrinos in many BSM scenarios. For instance, a light leptonic scalar $\phi$ can couple to neutrinos in the form of $\phi \nu \bar\nu$ or $\phi \nu \nu^c$ depending on the lepton number carried by $\phi$, which induces neutrino self-interactions and gives rise to interesting signals at both low-energy experiments~\cite{Pasquini:2015fjv, Berryman:2018ogk, Kelly:2019wow, Brdar:2020nbj,Deppisch:2020sqh} and high-energy colliders~\cite{deGouvea:2019qaz,Dev:2021axj,Agashe:2024owh}, as well as from astrophysical~\cite{Kolb:1987qy,Ng:2014pca, Shoemaker:2015qul, Heurtier:2016otg, Das:2017iuj, Kelly:2018tyg, Shalgar:2019rqe, Bustamante:2020mep, Esteban:2021tub,Chang:2022aas,Fiorillo:2022cdq,Fiorillo:2023ytr,Fiorillo:2023cas,Doring:2023vmk,Telalovic:2024cot} and cosmological~\cite{Cyr-Racine:2013jua, Archidiacono:2013dua, Huang:2017egl, Escudero:2019gvw,Barenboim:2019tux, Blinov:2019gcj,  Lyu:2020lps, Das:2020xke,Camarena:2024zck} observables; see Ref.~\cite{Berryman:2022hds} for a recent review. Another example of neutrinophilic mediators is the so-called Majoron particle $J$ with
interaction structure $J \bar\nu i\gamma_5 \nu$ originating from global symmetry breaking in seesaw models~\cite{Chikashige:1980ui, Gelmini:1980re, Schechter:1981cv}. In some scenarios, the scalar might also couple to neutrino and dark (matter) particle $\chi$, e.g., in the form of $\phi \bar\nu \chi$~\cite{Boehm:2003hm,Barranco:2010xt,Farzan:2012ev,Olivares-DelCampo:2017feq,Ghosh:2017jdy, Hagedorn:2018spx,Okawa:2020jea,Hufnagel:2021pso,Alvarado:2021fbw,Iguro:2022tmr,Herms:2023cyy, Higuchi:2023kbt,nu_DM:2024}. Such couplings may contribute to neutrino self-interactions at the 1-loop level~\cite{Bischer:2018zbd} or radiatively generate nontrivial electromagnetic properties of neutrinos. There are also some seesaw models with a scalar coupling to the active neutrinos and heavy neutrino $N$ via $\phi \bar{N} \nu$~\cite{Tao:1996vb,Ma:1998dn,Ma:2006km,Boehm:2006mi,Hambye:2009pw,Farzan:2010mr,Ibarra:2011gn,Farzan:2012sa,Farzan:2014gza,Li:2020pfy,Chianese:2020khl,Ge:2020jfn,Ge:2021snv,Chianese:2021toe}.

\begin{figure}[t!]
    \centering
    \includegraphics[width=0.165\textwidth]{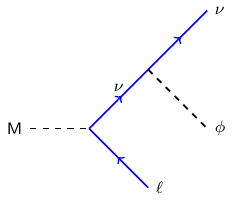}
    \includegraphics[width=0.125\textwidth]{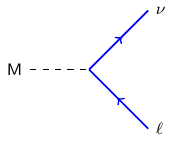}
    \includegraphics[width=0.165\textwidth]{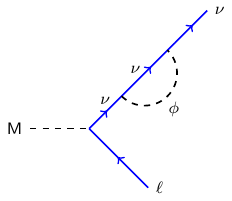} \\
    (a) \hspace{2cm} (b) \hspace{2cm} (c)
    \vspace{-5pt}
    \caption{Feynman diagrams for (a) the meson decay ${\sf M}^\pm \to \ell^\pm + \barparena{\nu} + \phi$, (b) the SM process ${\sf M}^\pm \to \ell^\pm + \barparena{\nu}$, and (c) the 1-loop correction due to the interaction in Eq.~(\ref{eqn:coupling:meson}). }
    \label{fig:diagram-meson}
    \vspace{-12.5pt}
\end{figure}

Such nonstandard interactions of neutrinos via neutrinophilic mediators induce new decay modes of SM particles, e.g., $\pi^\pm, K^\pm \to \ell^\pm + \barparena{\nu} / \chi + \phi$ (see Fig.~\ref{fig:diagram-meson}a) and $Z \to \nu + \bar{\nu} / \chi + \phi$~\cite{Barger:1981vd,Lessa:2007up,Berryman:2018ogk}.
Then the corresponding experimental data, e.g., the decay widths and the spectra of charged leptons $\ell^\pm$ from meson decays can be used to set limits on these decay rates as a function of the mediator mass $m_\phi$, as done by the PIENU~\cite{PIENU:2021clt} and NA62~\cite{NA62:2021bji} experiments using charged pion and kaon decays, respectively.
However, it is a common lore that these decay channels are potentially subject to the infrared (IR) divergences; the corresponding partial widths approach infinity as $m_\phi \to 0$ (see e.g., Refs.~\cite{Berryman:2018ogk, Carlson:2012pc, deGouvea:2019qaz, Dutta:2021cip}). This is clearly unphysical.
We show that the IR divergence is removed by including the interference between the 1-loop contribution (Fig.~\ref{fig:diagram-meson}c) and the 2-body decay (Fig.~\ref{fig:diagram-meson}b). This is reminiscent of the infrared behavior of interacting quantum field theories featuring massless fields within the general context of scattering amplitudes and the $S$-matrix in gauge theories~\cite{Bloch:1937pw, Yennie:1961ad, Kinoshita:1962ur, Lee:1964is, Weinberg:1965nx,Chung:1965zza,Kibble:1968lka, Kulish:1970ut,Grammer:1973db, Furugori:2020vdl, Hirai:2022yqw} (for reviews, see e.g., Refs.~\cite{Peskin:1995ev, Agarwal:2021ais}). In fact, this is expected as a natural consequence of the Kinoshita-Lee-Nauenberg (KLN) theorem~\cite{Kinoshita:1962ur, Lee:1964is}.

For illustration purposes, we focus on the following decay processes in this letter: exotic charged-meson decays ${\sf M}^\pm \to \ell^\pm + \barparena{\nu} + \phi$ with ${\sf M} = \pi,K$ and $\ell = e,\mu$, hadronic tau decays $\tau^\pm \to \pi^\pm + \barparena{\nu} + \phi$, and $Z$ boson decays $Z \to \nu + \bar{\nu} + \phi$.
More general cases involving dark matter $\chi$ or heavy neutrino $N$ in the final state such as ${\sf M}^\pm \to \ell^\pm + \chi/N + \phi$ with nonzero mass $m_{\chi / N}$ are also of great interest, e.g., for DM phenomenology and heavy neutrino searches, and will be reported in our forthcoming work~\cite{nu_DM:2024}.

We point out that summing up the tree and 1-loop contributions will not only give ``physical" constraints on the associated decays in the IR limit of small $\phi$ mass but also, in general, improve the constraints at large $\phi$ mass. When the mediator is heavy, the tree-level process is kinematically suppressed or forbidden and the BSM effects are dominated by the virtual mediator in the loop. This has far-reaching implications for the experimental limits on $m_{\phi}$ and its couplings. Currently, the major limiting factor for our analysis is the uncertainty on the determination of the meson decay constant (see Appendix~\ref{app:limits}), which we hope will be improved in the near future.


\begin{figure*}[!ht]
    \centering
    \includegraphics[height=0.325\textwidth]{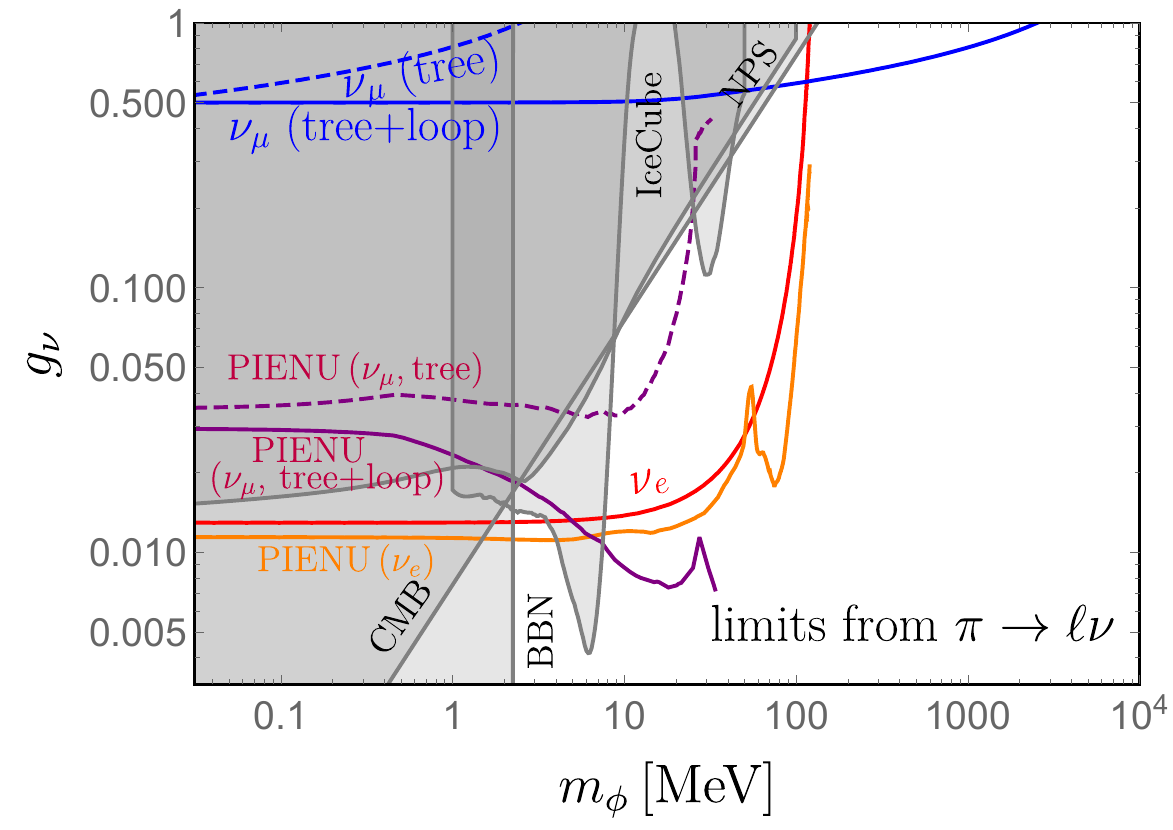}
    \includegraphics[height=0.325\textwidth]{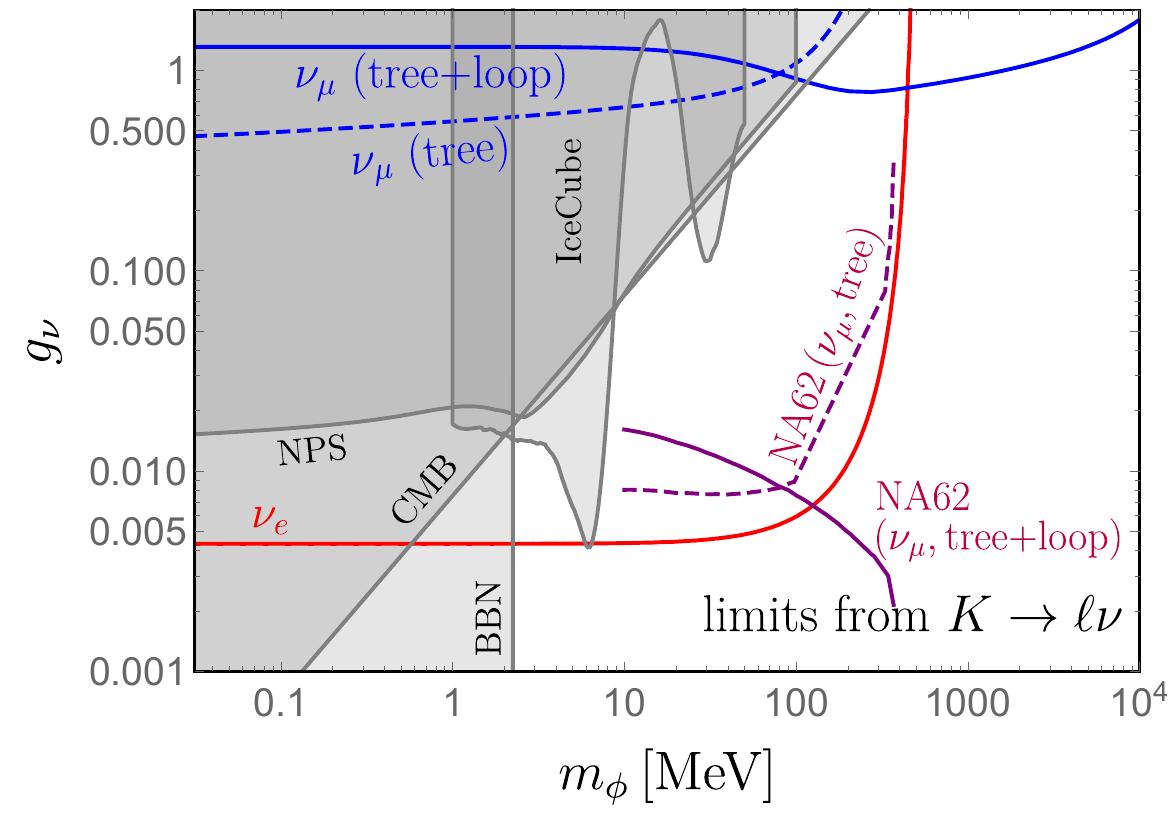}
    \vspace{-5pt}
    \caption{Limits on meson decays $\pi^\pm,\, K^\pm \to e^\pm/\mu^\pm + \barparena{\nu}+\phi$  at 90\% C.L., in the plane of $m_\phi$ and $g_\nu$. The dashed lines are the limits with only the tree-level contribution, while the solid lines are for those including the 1-loop contributions (cf.~Fig.~\ref{fig:diagram-meson}). The red and blue lines are respectively for the $g_{\nu_e}$ and $g_{\nu_\mu}$ couplings. The purple and orange lines are respectively the corresponding limits from PIENU (left)~\cite{PIENU:2021clt} and NA62 (right)~\cite{NA62:2021bji}. The gray-shaded regions are excluded by other constraints~\cite{Berryman:2022hds}. See the text for more details. }
    \label{fig:meson-nu-decays}
    \vspace{-10pt}
\end{figure*}

\section{Meson decays}

Let us first consider the meson decays ${\sf M}^\pm \to \ell^\pm + \barparena{\nu} + \phi$  with the light scalar $\phi$ (carrying no lepton number) emitted from the neutrino line (Fig.~\ref{fig:diagram-meson}a).  We consider the generic effective coupling of the form
\begin{equation}
\label{eqn:coupling:meson}
{\cal L} = g_\nu \phi \bar{\nu} \nu \,.
\end{equation}
The couplings of $\phi$ can be either flavor-diagonal or flavor-off-diagonal. It is well known that the partial widths for the 2-body leptonic decays ${\sf M}^\pm \to \ell^\pm + \barparena{\nu}$ are helicity-suppressed in the SM, i.e.,~proportional to the charged-lepton mass squared $m_\ell^2$. This is not the case for the 3-body decay; see Appendix~\ref{app:meson} for details. In the small $m_\phi$ limit, the partial width $\Gamma ({\sf M}^\pm \to \ell^\pm + \barparena{\nu} + \phi)$ can be written as
\begin{equation}
\label{eqn:width:meson}
\Gamma \simeq \frac{G_F^2 m_{\sf M}^3 f_{\sf M}^2 |V|^2 g_\nu^2}{128\pi^3} \left[ - x_{\ell {\sf M}} (1-x_{\ell {\sf M}})^2 \log x_{\phi {\sf M}} + C_2 (x_{\ell {\sf M}}) \right] \,,
\end{equation}
where $G_F$ is the Fermi constant, $m_{\sf M}$ and $f_{\sf M}$ are respectively the charged-meson mass and decay constant, $V$ is the CKM matrix element ($V_{ud}$ for pions and $V_{us}$ for kaons), $x_{ab} \equiv {m_a^2}/{m_b^2}$, and $C_2 (x_{\ell {\sf M}})$ is a dimensionless function of the mass ratio $x_{\ell {\sf M}}$, given in Eq.~(\ref{eqn:C2}) in Appendix~\ref{app:meson}.
It is apparent that the first term in Eq.~(\ref{eqn:width:meson}) is IR-divergent, i.e., goes to infinity in the limit of $m_\phi \to 0$, or equivalently $x_{\phi {\sf M}} \to 0$. For concreteness, we neglect the neutrino mass, which will not affect our results in the $\phi$ mass range of interest here.

The coupling in Eq.~(\ref{eqn:coupling:meson}) also induces a self-energy correction to the neutrino line, as shown in  Fig.~\ref{fig:diagram-meson}c. The amplitude ${\cal M}^{(0)}$ of the tree-level SM decay ${\sf M}^\pm \to \ell^\pm + \barparena{\nu}$, shown in Fig.~\ref{fig:diagram-meson}b, interferes with the amplitude ${\cal M}^{(1)}$ for the 1-loop diagram, when we calculate the partial width $\Delta \Gamma^{\rm loop} ({\sf M}^\pm \to \ell^\pm + \barparena{\nu})$. In particular, the interference term
\begin{equation}
{\rm Re} \left[ {\cal M}^{(0)\ast} {\cal M}^{(1)} \right] \propto g_\nu^2 \,,
\end{equation}
which is at the same order in $g_\nu$ as the partial width in Eq.~(\ref{eqn:width:meson}). The full expression for $\Delta \Gamma^{\rm loop} ({\sf M}^\pm \to \ell^\pm + \barparena{\nu})$ is given in Eq.~(\ref{eqn:width:meson:loop}) in Appendix~\ref{app:meson}. The IR-divergent part is $+x_{\ell {\sf M}} (1-x_{\ell {\sf M}})^2 \log x_{\phi {\sf M}}$ that exactly cancels out the first term of Eq.~(\ref{eqn:width:meson}), as expected from KLN theorem.

Taking into account the loop effects, we show in Fig.~\ref{fig:meson-nu-decays} the updated limits from $\pi^\pm$ and $K^\pm$ decays in the left and right panels, respectively.
We first report conservative limits, taking 90\% C.L. uncertainty ranges of the partial widths based on the information in the PDG data~\cite{ParticleDataGroup:2024cfk}; more details can be found in Appendix~\ref{app:limits}. The blue and red lines are for the muon and electron decay modes, respectively.

In the case of $\ell = e$, the logarithmic divergence is heavily suppressed by $x_{e {\sf M}} \equiv m_e^2 / m_{\pi,\,K}^2$, hence the $C_2$ term becomes more important in Eq.~(\ref{eqn:width:meson}) for $m_\phi \gtrsim$ eV. In other words, the IR divergence does not dominate the decay rate of ${\sf M}^\pm \to e^\pm + \barparena{\nu} + \phi$. The limits on ${\sf M}^\pm \to e^\pm + \barparena{\nu} + \phi$ from the ${\sf M}^\pm \to e^\pm + \barparena{\nu}$ data (red lines) are flat even in the limit of $m_\phi \to 0$, showing little differences between the tree-level and tree+loop-level limits, as expected.
When $m_\phi \ll m_{\sf M}$, the pion and kaon decay limits on $g_\nu$ in the electron channel are respectively $0.013$ and $4.3 \times 10^{-3}$.

In contrast, in the muon case $\ell = \mu$, as $m_\mu$ is comparable to $m_{\pi, K}$, the divergent behavior is noticeable in the $m_\phi \to 0$ limit, which is clear from the dashed blue lines in Fig.~\ref{fig:meson-nu-decays} (the label ``tree'').  The corresponding IR-free limits with the loop corrections are presented by the solid blue lines, labeled as ``tree+loop''. Notably, while the tree-level decay ${\sf M}^\pm \to \ell^\pm + \barparena{\nu} + \phi$ is kinematically forbidden when $m_\phi \geq m_{\sf M} - m_\ell$, the 1-loop contribution to the decay ${\sf M}^\pm \to \ell^\pm + \barparena{\nu}$ still exists. Consequently, the solid blue lines in Fig.~\ref{fig:meson-nu-decays} can extend to large $m_\phi$, even beyond the parent particle mass $m_{\sf M}$, whereas the dashed blue lines quickly vanish as $m_\phi$ gets closer to $m_{\sf M} - m_\mu$. With the loop contributions included, $\pi^\pm \to \mu^\pm + \barparena{\nu} + \phi$ gives $g_\nu < 0.50$ in the massless $\phi$ limit. For heavy $\phi$, this decay constrains $m_\phi$ up to $\sim 2.5$ GeV for $g_{\nu} <1$. The limit gets weaker in the large $m_\phi$ limit, as expected by the decoupling theorem.
The IR limit for the kaon decay is relatively weaker, i.e., $g_\nu <1.3$ is allowed, while for the mass range of $69 \, {\rm MeV} < m_\phi < 1.65 \, {\rm GeV}$ the coupling $g_\nu<1$. The dip feature of the solid blue line at around 200 MeV in the right panel of Fig.~\ref{fig:meson-nu-decays} is due to the substantial cancellation of the tree and 1-loop contributions.

These limits can be further improved by a dedicated shape analysis of the final decay products. The PIENU experiment has provided limits on the branching ratio (BR) of $\pi^\pm \to e^\pm/\mu^\pm + \barparena{\nu} + X$ as a function of the invisible $X$ mass~\cite{PIENU:2021clt}. We reinterpret them with our updated partial width calculations in both electron and muon channels, which are shown respectively by the orange and purple lines in the left panel of Fig.~\ref{fig:meson-nu-decays}. In the electron channel, we find that the limit of $g_\nu$ is improved by $\sim 12\%$ for small $m_\phi$. In contrast, the loop-included result (solid) differs from the tree-level one (dashed) in the muon channel. Qualitatively, when $m_\phi$ approaches the kinematic threshold, $m_\pi - m_\mu \simeq 34$ MeV, the tree-level contribution is highly suppressed by phase space and the loop contribution becomes the dominant BSM effect.
Therefore, the ``tree+loop'' limit gets much stronger at $m_\phi \gtrsim 10$ MeV. For $m_\phi \to 0$, the PIENU limit on $g_{\nu_\mu}$ is 0.029, while at $m_\phi \sim 10$ MeV, it is improved to $7.2\times 10^{-3}$.

Similarly, the NA62 experiment has reported the shape-analysis-based limits on ${\rm BR} (K^\pm \to \mu^\pm + \barparena{\nu} + X)$ with $X$ being a scalar~\cite{NA62:2021bji}. They are of the order of ${\cal O} (10^{-6})$ for $10\; {\rm MeV} < m_X < 370$ MeV, roughly three orders of magnitude stronger than the limits from the partial widths above.
The resultant ``tree'' and ``tree+loop'' limits are shown respectively by the dashed and solid purple lines in the right panel of Fig.~\ref{fig:meson-nu-decays}.
Again, with the 1-loop contribution included, the NA62 limits get much stronger, especially when $m_\phi$ is close to the threshold; at $m_\phi = 370$ MeV, the limit can reach down to $2.1\times 10^{-3}$.

We note that due to the existence of an off-shell neutrino propagator in the 3-body decay, the energy/angular distribution of the charged lepton from meson decays might be (mildly) affected, and the corresponding limits should be interpreted accordingly; see e.g., Fig.~7 of Ref.~\cite{NA62:2021bji}. Therefore, more dedicated analyses of the PIENU and NA62 data may improve to some extent the limits on $m_\phi$ and $g_{\nu}$ reported here.
We will examine this aspect in future work.

Speaking of the existing limits, all the gray-shaded regions in both panels of Fig.~\ref{fig:meson-nu-decays} show the exclusions by current terrestrial, astrophysical, and cosmological data~\cite{Berryman:2022hds}, i.e., those from the cosmic microwave background (CMB)~\cite{Archidiacono:2013dua,Escudero:2019gvw,Camarena:2024zck}, big bang nucleosynthesis (BBN)~\cite{Blinov:2019gcj},
IceCube High Energy Starting Events~\cite{Bustamante:2020mep}, and the high-energy neutrinos detected by IceCube from the neutrino point sources (NPSs) NGC 1068 and TXS 0506+056~\cite{Kelly:2018tyg,Doring:2023vmk}. Other existing limits, e.g., those from double-beta decays~\cite{Deppisch:2020sqh} (see also Ref.~\cite{Brune:2018sab}), SN1987A~\cite{Kolb:1987qy,Shalgar:2019rqe,Chang:2022aas, Fiorillo:2022cdq,Fiorillo:2023ytr,Fiorillo:2023cas,Telalovic:2024cot} and stellar cooling~\cite{Escudero:2019gzq}, are relatively weaker for the parameter space of our interest and hence not shown in Fig.~\ref{fig:meson-nu-decays}. More details of the limits are given in Appendix~\ref{app:limits}. Moreover, the coupling in Eq.~(\ref{eqn:coupling:meson}) induces 1-loop couplings of $\phi$ to the quarks and charged leptons~\cite{Chikashige:1980ui}, which would give additional limits from neutrino-electron and neutrino-nucleus scattering~\cite{Laha:2013xua}, e.g., those from Borexino~\cite{Borexino:2008gab} and COHERENT~\cite{COHERENT:2017ipa}. However, they are highly suppressed by the loop factor and the heavy $W$ and $Z$ particles in the loop, and are therefore not shown here.

Finally, as natural extensions, we have also calculated the partial widths for other cases and the corresponding meson decay limits:
({\it i}) The scalar couples to the charged leptons, i.e., $g_\ell \phi \bar{\ell} \ell$ (with $\ell = e,\mu$)~\cite{Carlson:2012pc, Krnjaic:2019rsv, Dutta:2021cip}, where the same cancellation happens. However, such couplings contribute to the anomalous magnetic moments of electron~\cite{Fan:2022eto} and muon~\cite{Muong-2:2023cdq}, which give rise to more stringent limits than the meson decay limits under consideration~\cite{Lindner:2016bgg}. Therefore we do not pursue this case further.
({\it ii}) For the pseudoscalar couplings to neutrinos and charged leptons, there is no IR divergence~\cite{Rai:2021vvq}.
({\it iii}) The analysis above can also be applied to the charged $D$ meson decays, i.e., $D^\pm \to \ell^\pm + \barparena{\nu} + \phi$, and also to the semileptonic $B$-meson decays. However, the corresponding limits are weaker than the ones shown here~\cite{Berryman:2018ogk, deGouvea:2019qaz}.
({\it iv}) If the scalar $\phi$ is replaced by a vector boson $Z'$, the corresponding partial width $\Gamma ({\sf M}^\pm \to \ell^\pm + \barparena{\nu} + Z')$ is dominated by the term $m_{\sf M}^4/m_{Z'}^2$ originating from the longitudinal polarization of $Z'$; this is much larger than the IR divergent term $m_\ell^2 \log ( m_{Z'}^2/ m_{\sf M}^2)$, see e.g., Refs.~\cite{Barger:2011mt,Carlson:2012pc,Laha:2013xua,Bakhti:2017jhm,Dutta:2021cip}.
We delve into this intriguing case in forthcoming work~\cite{nu_DM:2024}.


\begin{figure}[t]
    \centering
    \includegraphics[width=0.45\textwidth]{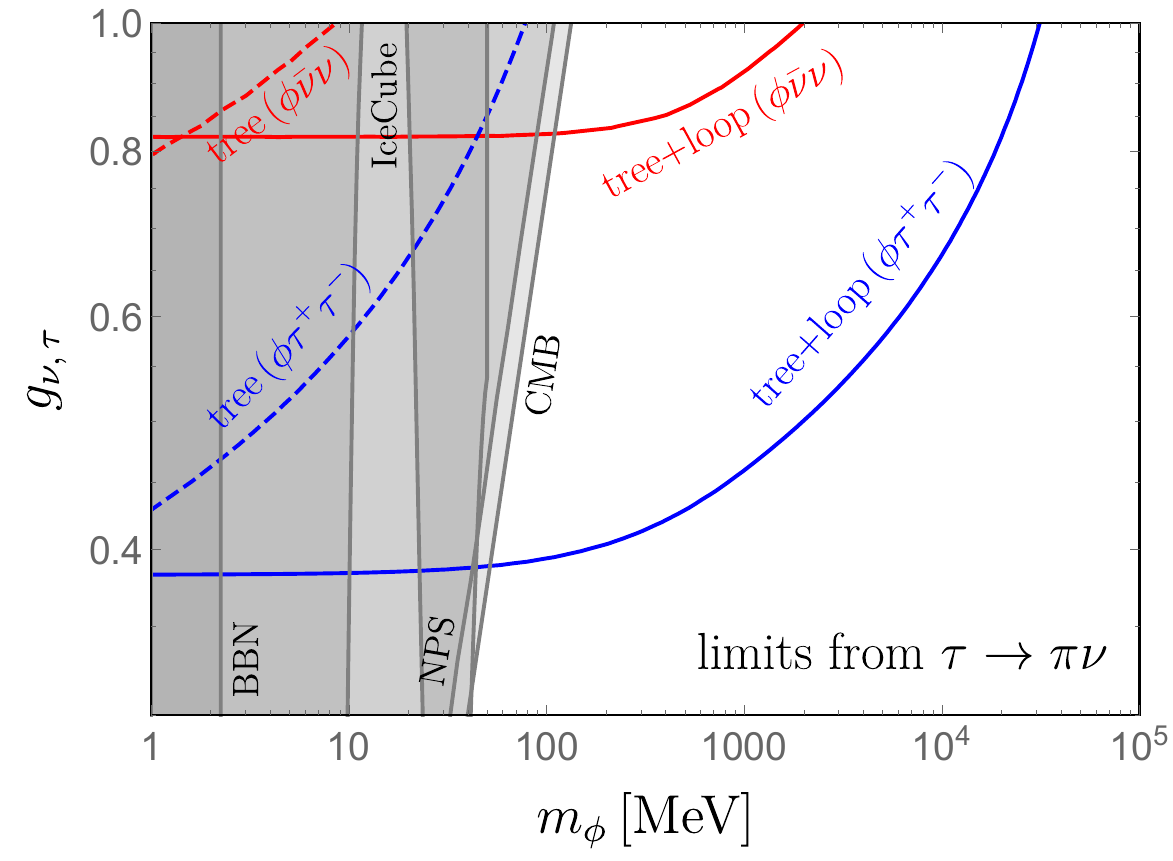}
    \vspace{-5pt}
    \caption{Limits on tau decay $\tau^\pm \to \pi^\pm + \barparena{\nu} + \phi$ at  90\% C.L. in the plane of $m_\phi$ and $g_{\nu,\tau}$. The red and blue lines are respectively for $g_\nu$ and $g_\tau$. The limits in gray are relevant only to $g_\nu$. The other notations are the same as in Fig.~\ref{fig:meson-nu-decays}. }
    \label{fig:tau}
    \vspace{-10pt}
\end{figure}

\section{Tau decays}

One of the dominant tau-lepton decay channels in the SM is $\tau^\pm \to \pi^\pm + \barparena{\nu}$, which is closely related to the charged-pion decays $\pi^\pm \to \ell^\pm + \barparena{\nu}$. Calculations of the decay channel $\tau^\pm \to \pi^\pm + \barparena{\nu} + \phi$ are similar to those for the charged-meson decays, and the details are given in Appendix~\ref{app:tau}. The resultant ``tree'' and ``tree+loop'' limits on $m_\phi$ and $g_\nu$ estimated with the $\tau^\pm \to\pi^\pm+ \barparena{\nu}$ partial width measurement are presented in Fig.~\ref{fig:tau} respectively by the dashed and solid red lines.
As in Fig.~\ref{fig:meson-nu-decays}, the gray shaded regions are excluded by current limits~\cite{Berryman:2022hds}.

Here we also consider the case of $\phi$ coupling to $\tau$ via $g_\tau \phi \tau^+ \tau^-$. The corresponding limits from $\tau^\pm \to \pi^\pm + \barparena{\nu}$ are shown by the (dashed) blue lines. The existing limits on $g_\tau$ are much weaker, mainly from the measurement of the anomalous $\tau$ magnetic moment. The current ATLAS constraint of $-0.057 < a_\tau < 0.024$~\cite{ATLAS:2022ryk} leads to the exclusion bound of $g_\tau > 1.1$~\cite{Lindner:2016bgg}, and is out of the presentation range in Fig.~\ref{fig:tau}.

We find that, once the 1-loop contribution is included, the allowed values of $g_{\nu,\tau}$ are smaller than respectively 0.82 and 0.38 in the $m_\phi \to 0$ limit. For $g_{\nu,\tau}<1$, $m_\phi$ is constrained up to 2.0 GeV and 31 GeV, respectively, which are well beyond the $\tau$ mass and the existing limits.

The pure leptonic decay channel $\tau^\pm \to \ell^\pm + \nu_\ell + \nu_\tau$ (with $\ell = e, \mu$) can also be used to set limits on exotic decays, i.e., $\tau^\pm \to \ell^\pm + \nu_\ell + \nu_\tau + \phi$ with $\phi$ emitted from the neutrino or charged-lepton lines~\cite{Lessa:2007up,Brdar:2020nbj}. However, with respect to the 3-body decay $\tau^\pm \to \pi^\pm + \barparena{\nu} + \phi$ considered above, these 4-body leptonic decays are both phase-space and BR-suppressed. Similarly, we expect that the 4-body decay channels from muon, i.e., $\mu^\pm \to e^\pm + \nu_e + \nu_\mu  + \phi$, will not give competitive limits.
Nevertheless, we will examine these 4-body decays in future work for completeness.


\section{$Z$ boson decays}

\begin{figure}[!t]
    \centering
    \includegraphics[width=0.45\textwidth]{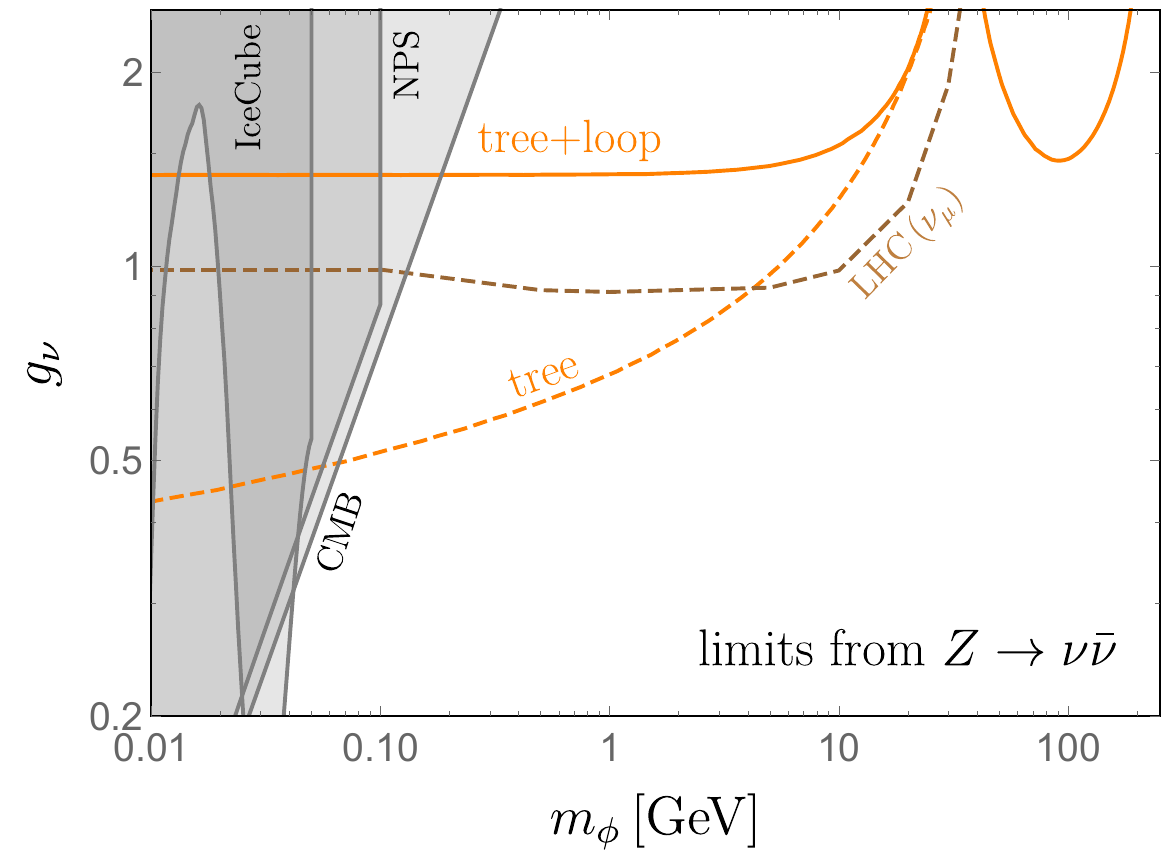}
    \vspace{-5pt}
    \caption{Limits on the decay
    $Z \to \nu + \bar\nu + \phi$ (orange) at 90\% C.L. in the plane of $m_\phi$ and $g_\nu$.  The brown line shows the LHC Run-3 prospect (for $\nu_\mu$). The other notations are the same as in Fig.~\ref{fig:meson-nu-decays}. }
    \label{fig:Z-med}
    \vspace{-10pt}
\end{figure}

The invisible $Z$ decay data can be utilized to set limits on our neutrinophilic mediator $\phi$ through the $Z \to \nu +\bar\nu + \phi$ decay channel.
The calculational details are given in Appendix~\ref{app:Z}. Just like in our previous cases, the tree-level contribution shows IR divergence which is removed by including the 1-loop contributions. But here the 1-loop corrections come from the neutrino self-energy, as well as from the $Z \nu \bar{\nu}$ vertex, unlike the meson and tau cases above. While we observe this cancellation even with massless neutrinos, it is interesting to compare our findings with the results in Ref.~\cite{Brdar:2020nbj}, where the IR divergence was regulated by the neutrino mass, which becomes relevant only in the regime $m_\phi\lesssim m_\nu$, far below the mass range of our interest here. Moreover, the difference in the IR behavior is presumably due to their different charge assignment for the scalar.
The orange lines in Fig.~\ref{fig:Z-med} show the resulting limits from invisible $Z$ data; $g_\nu<1.4$ is constrained for $m_\phi \ll m_Z$ with 1-loop contributions included. Due to the cancellation of the tree and loop contributions, the ``tree+loop'' limit gets much weaker at around $m_\phi \sim 30$ GeV (as shown by the ``gap'').

The current limits are indicated by the gray shaded region in Fig.~\ref{fig:Z-med}~\cite{Berryman:2022hds}. The search prospect of $\phi$ at the large hadron collider (LHC) Run-3 in the $W^\pm\to \mu^\pm+{\rm MET}$ channel with an integrated luminosity of 300 fb$^{-1}$ and 0.1\% systematics is shown by the brown line~\cite{Agashe:2024owh}. It is clear from the solid orange lines in Fig.~\ref{fig:Z-med} that, for the $\phi$ mediator, the invisible $Z$-decay data have excluded a sizable range of parameter space
beyond the current limits, complementing the prospect at the LHC.

One can also derive limits from the exotic $W$ boson decays, i.e., $W^\pm \to \ell^\pm + \barparena{\nu} + \phi$, and the calculations are very similar to the $Z$ boson case. However, the uncertainty $\Delta {\rm BR} (W^\pm \to \ell^\pm + \barparena{\nu}) \simeq 3.6 \times 10^{-3}$, is much larger than that from the invisible $Z$ data, $7.3 \times 10^{-4}$~\cite{ParticleDataGroup:2024cfk}. The resulting exclusions on the coupling are $g_\nu > 1$ and are thus not shown in Fig.~\ref{fig:Z-med}.

The invisible $Z$ data can also be used for other rare $Z$ decay channels; e.g., $Z \to \nu +\bar\nu +Z'$ with a neutrinophilic vector mediator $Z'$. This certainly carries nontrivial physics implications and features in mitigating the associated IR divergence, which are quite different from the case of ${\sf M}^\pm \to \ell^\pm + \barparena{\nu} + Z'$. We will defer the detailed discussion of the vector case for future publication~\cite{nu_DM:2024}.


\section{Discussions and conclusions}

In this letter, we have studied the exotic decays of charged mesons, tau lepton, and $Z$ gauge boson in the presence of a (light) neutrinophilic scalar $\phi$. We particularly focused on the IR divergence arising in the $m_{\phi}\to 0$ limit which is shown to be removed with the 1-loop contributions included.
The methodology here can also be applied to other decay channels, e.g., $\pi^0 \to \gamma + \gamma + \phi$ with $\phi$ coupling to photons or the channel $\pi^\pm \to e^\pm + \barparena{\nu} + a$ with pseudoscalar $a$ coupling to the $W$ mediator or the valence quarks of $\pi^\pm$~\cite{Bardeen:1986yb,Altmannshofer:2019yji,Bauer:2021mvw}. One may also constrain the hadronic couplings of a (light) scalar from meson and tau decays, e.g., in the channel of $\tau^\pm \to \pi^\pm + \barparena{\nu} + \phi$ with $\phi$ coupling to $\pi^\pm$ instead of $\barparena{\nu}$ or $\tau^\pm$.

Several implications of the 1-loop corrections are worth mentioning.
({\it i}) The 1-loop contributions are important in not only removing the IR divergence but also, in general, bringing new limits in the region of parameter space that is kinematically ``forbidden'' to constrain at the tree level (see the solid lines in Figs.~\ref{fig:meson-nu-decays} through \ref{fig:Z-med}).
({\it ii}) When conducting similar phenomenological studies, one should carefully include loop contributions to perform the theory calculations more accurately and place experimental bounds more robustly without the unphysical IR divergence.
({\it iii}) Some past experimental limits should be revisited accordingly, e.g., the PIENU and NA62 limits re-interpreted in our study.

In this work, the effects of the scalar $\phi$ are either via the extra decay channels of the charged mesons, the tau lepton and $Z$ boson, or via virtual correction to the wavefunction renormalization of the neutrino field. In both cases,
the limits on the $\phi$ couplings to neutrinos derived here using the high-precision SM decay processes are complementary to the direct limits on the nonstandard interactions (NSIs) of neutrinos from SN1987A~\cite{Shalgar:2019rqe}, e.g., using $\phi$-induced neutrino self-scattering process $\nu \nu \to \nu \nu$. See Appendix~\ref{app:limits} for more details.

In summary, the SM should be IR-finite, as stated by the KLN theorem.
This holds even in the presence of BSM couplings. We have demonstrated this general feature with a scalar $\phi$ interacting with the active neutrinos and $\tau^\pm$. The loop contributions will not only be important for the IR divergence cancellation, but also have far-reaching implications for both phenomenological studies and the correct interpretation of experimental limits.

\bigskip

\acknowledgments

We thank Kaladi Babu, Bhaskar Dutta, Damiano Fiorillo, Sudip Jana, Kevin Kelly, Lorenzo Ricci, Oleksandr Tomalak, Xun-Jie Xu and Yue Zhang for useful discussions and comments on the draft. BD is supported by the U.S. Department of Energy grant No. DE-SC 0017987. The work of DK was supported in part by the
DOE Grant No. DE-SC0010813.  DS is supported by NSF Grant No. PHY-2210361 and by the Maryland Center for Fundamental Physics. KS is supported by the U.S. Department of Energy grant DE-SC0009956. YZ is supported by the National Natural Science Foundation of China under grant No. 12175039, the 2021 Jiangsu Shuangchuang (Mass Innovation and Entrepreneurship) Talent Program No. JSSCBS20210144, and the ``Fundamental Research Funds for the Central Universities''. BD, DK, and KS acknowledge the Center for Theoretical Underground Physics and Related Areas (CETUP* 2024) and the Institute for Underground Science at SURF for hospitality and for providing a stimulating environment, where this work was finalized.

\appendix

\begin{widetext}

\section{Details of the limits}
\label{app:limits}

Here we collect in Table~\ref{tab:data} all the data used in the Letter for the meson, tau, and $Z$ decay limits. For instance, from the $\pi \to e \nu$ data in the table, the $1\sigma$ range uncertainty is
\begin{equation}
\label{eqn:uncertainty}
\Delta \Gamma (\pi \to e\nu) =  \frac{{\rm BR} (\pi \to e \nu)}{\tau_{\pi^\pm}} \left[ \frac{\Delta \tau_{\pi^\pm}}{\tau_{\pi^\pm}} + \frac{\Delta {\rm BR} (\pi \to e \nu)}{{\rm BR} (\pi \to e \nu)} + \frac{2 \Delta f_{\pi^\pm}}{f_{\pi^\pm}}  \right] \,,
\end{equation}
where $\tau_{\pi^\pm}$ and ${\rm BR} (\pi \to e\nu)$ are the experimental values of the charged pion lifetime and the BR for the channel $\pi^\pm \to e^\pm \nu$~\cite{ParticleDataGroup:2024cfk}. $f_{\pi^\pm}$ is the charged pion decay constant, independently determined from the lattice calculations~\cite{fpi}. The factor of 2 for $f_{\pi^{\pm}}$ in Eq.~\eqref{eqn:uncertainty} is for the square dependence of $\Delta \Gamma (\pi \to e\nu)$ on the decay constant, cf.~Eq.~\eqref{eqn:width:meson}. It should be noted that in Eq.~(\ref{eqn:uncertainty}) the uncertainty  $\Delta \Gamma (\pi \to e\nu)$ is dominated by the decay constant, whose precision is at the order of ${\cal O} (10^{-2})$ (cf. Table~\ref{tab:data}). Other more precise measurements, such as the charged pion and kaon masses, and the Fermi constant, contribute much less to the uncertainty $\Delta \Gamma (\pi \to e\nu)$, and are therefore neglected in our calculations.  To apply the limits at the 90\% C.L., we multiply the $1\sigma$ uncertainties by a factor of 1.64.


Neutrino NSIs may potentially have impacts on the supernova core.
For sufficiently large NSIs, neutrinos will form a fluid in the supernova core,
and the outflow of neutrino fluid
streams as a fireball~\cite{Chang:2022aas}. Therefore, the impact of large neutrino NSIs on supernova dynamics is small~\cite{Fiorillo:2023ytr,Fiorillo:2023cas}.
There are also constraints from the scattering of the supernova
neutrinos with the cosmic neutrino background en route to Earth for $m_\phi \gtrsim$ keV. However, the corresponding limits are rather weak for the parameter space of $m_\phi$ and $g_\nu$ we are interested in~\cite{Kolb:1987qy,Shalgar:2019rqe}.

Regarding the double-beta decay limits, the limits from the neutrinoless double-beta decay searches, e.g., those from Ref.~\cite{Brune:2018sab}, are not applicable to the Dirac neutrino coupling in our Eq.~(\ref{eqn:coupling:meson}). Instead, the two-neutrino double-beta decays can be used to set limits on $m_\phi$ and $g_\nu$~\cite{Deppisch:2020sqh}. In particular, the scalar $\phi$ could induce the effective four-neutrino interaction $G_S(\bar\nu_e \nu_\alpha) (\bar\nu_e \nu_\beta)$, with $\alpha,\;\beta = e,\; \mu\; \tau$ the flavor indices, and the coefficient $G_S = g_\nu^2 / (t + m_\phi^2)$. It is expected that $t \sim p_F^2$, with $p_F \simeq 100$ MeV being the Fermi momentum. The most stringent double-beta decay limit $G_S < 3.2 \times 10^8 G_F$ implies a rather weak constraint on the coupling, i.e. $g_\nu \gtrsim 6.1$ for $m_\phi \lesssim p_F$~\cite{Deppisch:2020sqh}, and therefore, is not shown in our figures. For heavier $\phi$, the double-beta decay limit will be even weaker.

\begin{table*}[!t]
    \caption{\label{tab:data} Data used for the limits, with $\tau$ (or $\Gamma$) being the lifetime (or width) of the decaying parent particle, ${\rm BR}$ the corresponding BR for the channels in the first column, and $\Delta \Gamma$ the $1\sigma$ uncertainties for the partial widths. The last column is the experimental limits from the spectrum analysis of the charged leptons involved. Taken from PDG~\cite{ParticleDataGroup:2024cfk} unless otherwise specified.}
\begin{tabular}{cccc}
\hline\hline
Channel & Data & $1\sigma$ limits & Relevant expt. \\ \hline
\multirow{3}*{$\pi \to e \nu$} & $\tau_{} = (2.6033 \pm 0.0005) \times 10^{-8}$ sec & \multirow{3}*{$\Delta \Gamma =5.08\times10^{-20}$ MeV} & \multirow{3}*{PIENU~\cite{PIENU:2021clt}} \\
&  ${\rm BR} = (1.230 \pm 0.004) \times 10^{-4}$ & & \\
&  $f_{\pi^\pm} = (130.2 \pm 0.8)$ MeV~\cite{fpi} & & \\ \hline
\multirow{3}*{$\pi \to \mu \nu$} & $\tau_{} = (2.6033 \pm 0.0005) \times 10^{-8}$ sec & \multirow{3}*{$\Delta \Gamma =3.31\times10^{-16}$ MeV} & \multirow{3}*{PIENU~\cite{PIENU:2021clt}} \\
&  ${\rm BR} = 0.9998770 \pm 0.0000004$ & & \\
&  $f_{\pi^\pm} = (130.2 \pm 0.8)$ MeV~\cite{fpi} & & \\  \hline
\multirow{3}*{$K \to e \nu$} & $\tau_{} = (1.2380 \pm 0.0020) \times 10^{-8}$ sec &
\multirow{3}*{$\Delta \Gamma =1.90\times10^{-20}$ MeV} & \\
&  ${\rm BR} = (1.582 \pm 0.007) \times 10^{-5}$ & & \\
&  $f_{K^\pm} = (155.7 \pm 0.7)$ MeV~\cite{fpi} & & \\ \hline
\multirow{3}*{$K \to \mu \nu$} & $\tau_{} = (1.2380 \pm 0.0020) \times 10^{-8}$ sec & \multirow{3}*{$\Delta \Gamma =6.71\times10^{-16}$ MeV} & \multirow{3}*{NA62~\cite{NA62:2021bji}} \\
&  ${\rm BR} = 0.6356 \pm 0.0011$ & & \\
&  $f_{K^\pm} = (155.7 \pm 0.7)$ MeV~\cite{fpi} & & \\ \hline
\multirow{3}*{$\tau \to \pi \nu$} & $\tau_{} = (2.093 \pm 0.005) \times 10^{-13}$ sec & \multirow{3}*{$\Delta \Gamma =6.77\times10^{-12}$ MeV} & \\
&  ${\rm BR} = 0.1082 \pm 0.0005$ & & \\
&  $f_{\pi^\pm} = (130.2 \pm 0.8)$ MeV~\cite{fpi} & & \\ \hline
\multirow{2}*{$Z \to \nu \bar\nu$} & $\Gamma = (2.4955 \pm 0.0023)$ GeV & \multirow{2}*{$\Delta \Gamma =1.83$ MeV} & \\
&  ${\rm BR} = 0.20000 \pm 0.00055$ & & \\ \hline\hline
\end{tabular}
\end{table*}

\section{Meson decay ${\sf M} \to \ell + \nu + \phi$}
\label{app:meson}

For the meson decay ${\sf M} (p) \to \ell (p_\ell) + \nu (p_\nu) + \phi (p_\phi)$, with $\phi$ coupling to neutrinos with the strength $g_\nu$, the squared amplitude  is given by
\begin{equation}
\label{eqn:ampsq:pi1}
\sum |{\cal M} ({\sf M} \to \ell + \nu + \phi)|^2 =
\frac{8 g_\nu^2 f_{\sf M}^2 G_F^2 |V_{}|^2}{q^4}
\left\{ q^4 (p_\ell \cdot p_\nu) + m_\ell^2 \Big[ 2 (q\cdot p_\nu) (q^2 + (q\cdot p_\ell)) - q^2 (p_\ell \cdot p_\nu) \Big] \right\} \,,
\end{equation}
with $q$ being the momentum of the neutrino mediator. After a lengthy calculation, one can find the total partial width to be
\begin{equation}
\Gamma ( {\sf M} \to \ell + \nu + \phi ) = \frac{g_\nu^2 G_F^2 m_{\sf M}^3 f_{\sf M}^2 |V|^2}{128\pi^3} f_1 (x_{\phi {\sf M}},\,x_{\ell {\sf M}}) \,,
\end{equation}
with $x_{ab} \equiv m_a^2/m_b^2$, and the dimensionless function $f_1 (x_1,\,x_2)$ is given by
\begin{eqnarray}
f_1 (x_1,\,x_2) &=& \frac13 \frac{\lambda^{1/2} (1,x_1,x_2)}{1-x_2} \Big( 1 + 10 x_1 + x_1^2 - x_2 (9 - 6x_1 + x_1^2) + x_2^2 (18- 19x_1) - 10 x_2^3 \Big) \nonumber \\
&& + \Big( 2 x_1 (1+x_1) - x_2 \left(1-3x_1^2\right) - 2 x_2^2 (1-3 x_1) + x_2^3 \Big) {\rm arctanh} \frac{\lambda^{1/2} (1,x_1,x_2)}{1-x_1+x_2} \nonumber \\
&& - \left[  x_1 (1-3x_2^2) - x_2 (1-x_2)^2 + ( 2 - x_2  + 4 x_2^2 -3 x_2^3 ) \frac{x_1^2}{(1-x_2)^2} \right] {\rm arctanh} \frac{(1-x_2)\lambda^{1/2} (1,x_1,x_2)}{(1-x_2)^2 + x_1 (1+x_2)} \,,
\end{eqnarray}
where
\begin{equation}
\lambda (a,\, b,\, c) \equiv a^2 + b^2 + c^2 - 2ab - 2ac - 2 bc \,.
\end{equation}
For sufficiently small $x_1$,
\begin{eqnarray}
f_1 (x_1,\,x_2) &\simeq& - x_2 (1 + 2x_2 - x_2^2) {\rm arctanh} \frac{1-x_2}{1+x_2} \nonumber \\
&& + \frac16 (1-x_2) \Big[ 2 - 4 x_2 (4-5x_2) - 3 x_2 (1-x_2) \log \frac{x_1^2 x_2}{(1-x_2)^4} \Big] \,.
\end{eqnarray}
Then we get the IR-divergent part, i.e., the first term in Eq.~(\ref{eqn:width:meson}), and the rest of $f_1 (x_1,x_2)$ is the finite function
\begin{eqnarray}
\label{eqn:C2}
C_2 (x) &\simeq& - x (1 + 2x - x^2) {\rm arctanh} \frac{1-x}{1+x} \nonumber \\
&& + \frac16 (1-x) \Big[ 2 - 4 x (4-5x) - 3 x (1-x) \log \frac{x}{(1-x)^4} \Big] \,.
\end{eqnarray}

The self-energy of neutrino with momentum $p$ in $d$ dimensions due to its interaction with the scalar $\phi$ can be written as
\begin{equation}
\Sigma (p) = \slashed{p} \times \frac{i}{(4\pi)^2} \left( \frac{4\pi}{m_\phi^2} \right)^\varepsilon \frac{\Gamma (1+\varepsilon)}{\varepsilon} \frac{1}{2-\varepsilon} \,,
\end{equation}
where $\varepsilon = (4-d)/2$. The ultraviolet (UV) divergence in the limit of $\varepsilon \to 0$ can be removed by the standard procedure of adding appropriate counterterms in the $\overline{\text{MS}}$ scheme with the on-shell condition $p^2=0$ for the outgoing neutrino (we have set the neutrino mass to be zero). On the other hand, in the limit of $m_\phi \to 0$, there is an infrared divergence, which is in the form of $\log (m_\phi^2/m_{\sf M}^2)$. More explicitly, for interference term between Figs.~\ref{fig:diagram-meson} (b) and (c), its contribution to the width of the decay ${\sf M} \to \ell + \nu$ is
\begin{equation}
\label{eqn:width:meson:loop}
\Delta \Gamma^{\rm loop} ( {\sf M} \to \ell + \nu ) = - \frac{g_\nu^2 G_F^2 m_{\sf M} m_\ell^2 f_{\sf M}^2 |V|^2}{128\pi^3} \left( 1 - \frac{m_\ell^2}{m_{\sf M}^2} \right)^2 f_1^{\rm loop} (x_{\phi {\sf M}},\,x_{\ell {\sf M}}) \,,
\end{equation}
where the dimensionless function
\begin{equation}
f_1^{\rm loop} (x_1,\, x_2) = \frac52 - \log \frac{x_1 (1-x_2)^2}{16\pi^2}
\end{equation}
leads to the IR term proportional to $x_{\ell {\sf M}} (1-x_{\ell {\sf M}})^2 \log x_{\phi {\sf M}}$, which cancels exactly with the first term in Eq.~(\ref{eqn:width:meson}) from the 3-body decay ${\sf M} \to \ell + \nu + \phi$, as expected on general grounds from the KLN theorem. We emphasize that the IR and UV divergences are independent of each other, and  cancellation of the UV divergence by adding counterterms should not affect the IR behavior.

\section{Tau decay $\tau \to \pi + \nu + \phi$}
\label{app:tau}

In the SM, the matrix element for the semileptonic decay $\tau \to \pi + \nu$ is closely correlated with that for $\pi \to \ell + \nu$. For the case of $\phi$ coupling to the neutrino, we can easily obtain the squared amplitude for the decay $\tau (p) \to \pi (p_\pi) + \nu (p_\nu) + \phi (p_\phi)$:
\begin{equation}
\label{eqn:ampsq:tau}
\frac12 \sum |{\cal M}_\nu (\tau \to \pi + \nu + \phi) |^2 =
\frac{4 g_\nu^2 f_{\pi}^2 G_F^2 |V_{ud}|^2}{q^4}
\left\{ q^4 (p \cdot p_\nu) + m_\tau^2 \Big[ 2 (q\cdot p_\nu) ((q\cdot p)-q^2) - q^2 (p \cdot p_\nu) \Big] \right\} \,,
\end{equation}
where the factor of $1/2$ is for averaging over the spins of tau in the initial state. Then the partial width reads
\begin{equation}
\label{eqn:width:tau}
\Gamma ( \tau \to \pi + \nu + \phi ) = \frac{g_\nu^2 G_F^2 m_{\tau}^3 f_{\pi}^2 |V_{ud}|^2}{256\pi^3} f_2 (x_{\phi \tau},\, x_{\pi \tau}) \,,
\end{equation}
with the dimensionless function
\begin{eqnarray}
f_2 (x_1,\, x_2) &=& - \frac{\lambda^{1/2} (1,\,x_1,\,x_2)}{3} \left[ (10 + x_1^2-8 x_2+x_2^2) + \frac{x_1}{1-x_2} ( 19 - 6 x_2 - 10 x_2^2 ) \right] \nonumber \\
&& - \Big( 1 - x_2 (2+x_2) + 2 x_1 \left(3+ x_2^2\right) + x_1^2 (3+2 x_2)  \Big) {\rm arctanh} \frac{\lambda^{1/2} (1,\,x_1,\,x_2)}{1-x_1+x_2} \nonumber \\
&& + \left[ (1-x_2)^2 + 2x_1 (3-x_2^2) + \frac{x_1^2}{(1-x_2)^2} \Big( 3 - x_2 (4 - x_2 + 2x_2^2) \Big) \right] {\rm arctanh} \frac{(1-x_2)\lambda^{1/2} (1,x_1,x_2)}{(1-x_2)^2 + x_1 (1+x_2)}. \nonumber \\
&&
\end{eqnarray}
In the limit of $x_1 \to 0$, we have
\begin{equation}
f_2 (x_1,\, x_2) \simeq -\frac13 (1-x_2) (10-8x_2+x_2^2) - (1-x_2)^2 \log \frac{x_1}{(1-x_2)^2} - x_2^2 \log x_2 \,.
\end{equation}
The loop contribution to the decay $\tau \to \pi + \nu$ is
\begin{equation}
\label{eqn:width:loop:tau}
\Delta \Gamma^{\rm loop} ( \tau \to \pi + \nu ) = - \frac{g_\nu^2 G_F^2 m_{\tau}^3 f_{\pi}^2 |V_{ud}|^2}{256\pi^3} \left( 1 - \frac{m_\pi^2}{m_\tau^2} \right)^2 f_1^{\rm loop} (x_{\phi\tau},\,x_{\pi\tau}) \,.
\end{equation}

For the case of $\phi$ coupling to the tau with the strength $g_\tau$, the amplitude square is very similar to Eq.~(\ref{eqn:ampsq:tau}):
\begin{equation}
\frac12 \sum |{\cal M}_\tau (\tau \to \pi + \nu + \phi) |^2 =
\frac{4 g_\nu^2 f_{\pi}^2 G_F^2 |V_{ud}|^2}{(q^2-m_\tau^2)^2}
\left\{ q^4 (p \cdot p_\nu) + m_\tau^2 \Big[ 2 (q\cdot p_\nu) ((q\cdot p)+q^2) - q^2 (p \cdot p_\nu) \Big] \right\} \,,
\end{equation}
with $q$ being the momentum of the tau propagator. For the partial width, one only needs to replace the coupling $g_\nu$ by $g_\tau$ in Eq.~(\ref{eqn:width:tau}), and the corresponding dimensionless function is
\begin{eqnarray}
f_3 (x_1,\,x_2) &=& - \frac{\lambda^{1/2} (1,x_1,x_2)}{6} \Big[ 119 - 115 x_2 + 2 x_2^2 -x_1 ( 61 - 26x_2) + 2 x_1^2 \Big] \nonumber \\
&& +  \sqrt{x_1 (4-x_1)} (1-x_2) \Big( 2 (9-x_2) - x_1 (3+x_2) \Big) \left[ \frac{\pi}{2} + \arctan \frac{x_1 (3 - x_1 + x_2) \lambda^{-1/2} (1,x_1,x_2)}{\sqrt{x_1(4-x_1)}} \right] \nonumber \\
&& + \Big( 8 - 16 x_2 + 7x_2^2 -2 x_1 \left( 18 - 12x_2 + x_2^2 \right)  + x_1^2 \left(3-2 x_2\right) \Big) {\rm arctanh} \frac{\lambda^{1/2} (1,x_1,x_2)}{1-x_1+x_2} \nonumber \\
&& + x_2^2 (1-x_1)^2 {\rm arctanh} \frac{(1-x_1)\lambda^{1/2} (1,x_1,x_2)}{(1-x_1)^2 - x_2 (1+x_1)} \,.
\end{eqnarray}
In the limit of $x_1 \to 0$,
\begin{equation}
f_3 (x_1,\, x_2) \simeq -\frac16 (1-x_2) (119 -115 x_2 + 2x_2^2) - 4(1-x_2)^2 \log \frac{x_1}{(1-x_2)^2} - x_2^2 \log x_2 \,.
\end{equation}
The loop contribution is the same as in Eq.~\eqref{eqn:width:loop:tau} with $g_\nu$ replaced by $g_\tau$ and multiplied by a factor of 4.

\section{$Z$ boson decay $Z \to \nu + \bar\nu + \phi$}
\label{app:Z}

For the decay $Z (p) \to \nu (p_{\nu}) + \bar\nu (p_{\bar\nu}) + \phi (p_\phi)$, the amplitude square is
\begin{eqnarray}
\frac13 \sum |{\cal M} (Z \to \nu + \bar{\nu} + \phi)|^2 &=& \frac{2\sqrt2 g_\nu^2 m_Z^2 G_F}{3q^4}
\bigg[ 2 (q_1 \cdot p_{\nu}) (q_1 \cdot p_{\bar\nu}) + 2 (q_2 \cdot p_{\nu}) (q_2 \cdot p_{\bar\nu}) - (q_1^2+q_2^2) (p_{\nu} \cdot p_{\bar\nu}) \nonumber \\
&& + \frac{2}{m_{Z}^2} \Big( 2 (p \cdot p_{\bar\nu}) (p \cdot q_1) (q_1 \cdot p_{\nu}) + 2 (p \cdot p_{\nu}) (p \cdot q_2) (q_2 \cdot p_{\bar\nu}) - (q_1^2+q_2^2) (p \cdot p_{\nu}) (p \cdot p_{\bar\nu}) \Big) \bigg] \,, \nonumber \\ &&
\end{eqnarray}
where the factor of $1/3$ is for averaging the spins of $Z$ boson, and $q_{1,\,2} = p_\phi + p_{\nu,\, \bar\nu}$ is the momentum of the (anti)neutrino propagator. Then the corresponding partial decay width reads
\begin{equation}
\label{eqn:width:Z}
\Gamma (Z \to \nu \bar\nu \phi)  = \frac{g_\nu^2 G_F m_Z^3}{96 \sqrt2 \pi^3} f_{4} (x_{\phi Z}) \,,
\end{equation}
with the dimensionless function
\begin{equation}
f_{4} (x) = - \frac16 (1 - x) (17+8x-x^2) -(1+3x) \log x \,.
\end{equation}
The dimensionless function for the corresponding loop contribution to $Z \to \nu \bar\nu$ is, with the prefactors the same as in Eq.~(\ref{eqn:width:Z}):
\begin{eqnarray}
f_2^{\rm loop} (x) &=& -\frac32 + \log x + 2(1 - \log x) - \frac{2+3x+2x \log x}{(1+x)} - x (3+ 2 \log x) \log \left( \frac{x}{1+x} \right) \nonumber \\
&& + \frac{1}{1+x} {_2} f_1^{(0,0,1,0)} \left( 1,1,3, -\frac{1}{x} \right)
+ \frac{2}{x} {_2} f_1^{(0,0,1,0)} \left( 1,2,3, -\frac{1}{x} \right) + \frac{1}{x} {_2} f_1^{(0,1,0,0)} \left( 1,2,3, -\frac{1}{x} \right) \nonumber \\
&& + 2 \int_0^1 {\rm d}b \left[
{_2} F_1^{(0,1,0,0)} \left( 1,0,2, \frac{b}{x(1-b)} \right) - {_2} F_1^{(0,0,1,0)} \left( 1,0,2, \frac{b}{x(1-b)} \right) \right] \,,
\end{eqnarray}
where ${_2} f_1$ and ${_2} F_1$ are respectively the hyper-geometric and regularized hyper-geometric functions.

\end{widetext}

\bibliographystyle{JHEP}
\bibliography{updated_limits_refs}

\end{document}